\documentclass{article}
\usepackage{graphicx} 

\usepackage{amsmath}
\usepackage{amsfonts}
\usepackage{amssymb}
\date{\today}
\usepackage{float}

\newcommand{\ee}{\end{equation}}
\newcommand{\eea}{\end{eqnarray}}
\newcommand{\be}{\begin{equation}}
\newcommand{\bea}{\begin{eqnarray}}

\newcommand{\insertplot}[5]{\begin{figure}
 \hfill\hbox to 0.05in{\vbox to #5in{\vfill
 \inputplot{#1}{#4}{#5}}\hfill}
 \hfill\vspace{-.1in}
 \caption{#2}\label{#3}
 \end{figure}}
 \newcommand{\inputplot}[3]{
 \special{ps: plotfile #1}
}
\newcounter{fig}

\begin{document}

\title{Frozen states of charged boson stars}

\date{\today}

\author{
{\large Yves Brihaye}$^{1}$, and
{\large Betti Hartmann} $^{2}$
\\
\\
$^{1}${\small Physique de l'Univers, Universit\'e de
Mons, 7000 Mons, Belgium}
\\
$^{2}${\small Department of Mathematics, University College London, Gower Street, London, WC1E 6BT, UK}
}
\maketitle

\begin{abstract}
In this paper, we study frozen states of charged boson stars. These solutions
are globally regular and exist in a U(1) gauged scalar field model minimally coupled to gravity for suitable choices of the coupling constants. These configurations are field theoretical realizations of the Mazur-Mottola solution with a de Sitter interior, a black hole exterior and a thin shell that interpolates between the two and replaces the event horizon. We demonstrate that standard electrodynamics is sufficient to find these frozen states, but that the self-interaction of the scalar field is crucial.
Adding Horndeski vector-tensor gravity to the model allows the frozen states to exist without self-interaction though. The frozen states possess one stable and one unstable lightring, the former inside the thin shell, the latter in the black hole exterior.
\end{abstract}

\section{Introduction}

General Relativity is {\it per se} a deterministic theory, i.e. given an initial data set, the entire future and past should be given by the Einstein equation. However, determinism
breaks down at space-time singularities. The weak cosmic censorship conjecture \cite{Penrose:1969pc} suggests that these space-time singularities should always be hidden
behind an event horizon, i.e. exist at the center of black holes, and hence be non-observable from outside the horizon. Nevertheless, they pose a general problem for the deterministic nature of the theory of General Relativity. Hence, {\it black hole mimickers}
are an active area of research. These mimickers are (hypothetical) objects that look very much like black holes from afar, but do not possess an event horizon and - more importantly in the context of the discussion above - no physical singularity at the center (for a recent review see e.g. \cite{Bambi:2025wjx}). Boson stars are such a viable mimicker 
\cite{Kaup:1968zz,Schunck:1999pm, Friedberg:1986tp,Jetzer:1991jr, Schunck:2003kk, 
Liebling:2012fv}. Another possibility are regular black holes, i.e. black holes
that have a horizon, but no space-time singularity. The first solution of this type was presented by Bardeen \cite{bardeen} and was consequently interpreted to have as its source a magnetic monopole within the framework of non-linear electrodynamics \cite{Ayon-Beato:1998hmi, Ayon-Beato:2000mjt}. In contrast to regular black holes, frozen stars do not possess an event horizon, but a boundary with finite thickness that matches a non-singular interior to a black hole exterior. The first solution of this type has been given in \cite{Mazur:2001fv}. This is a compact object made of a perfect fluid with equation of state $\rho=-p$ (i.e. corresponding to de Sitter) in its interior and
a Schwarzschild solution in its exterior. The interior and exterior are matched across a shell of finite thickness which replaces the event horizon. A concrete realization motivated from String Theory has been discussed in \cite{Brustein:2021lnr, Brustein:2023cvf}.
Recent studies \cite{Zhao:2025hdg, Wang:2023tdz,Yue:2023sep} have extended these ideas by introducing a complex, ungauged scalar field in the original Bardeen model, leading to the construction of boson star solutions with nonlinear electrodynamics. Although horizonless, these configurations can develop a quasi-horizon at an intermediate radius, while maintaining regularity at the core. 

In this paper, we will demonstrate that non-linear electrodynamics is not necessary to find these frozen stars. In the first part of this paper, we will use a U(1) gauged scalar field
model minimally coupled to gravity. This has been used previously to study 
spherically symmetric, static and charged black holes that carry a cloud, a so-called Q-cloud, of scalar fields on the horizon \cite{Herdeiro:2020xmb, Hong:2020miv}. In particular, it was shown that both the presence of the electric field as well as the self-interaction are necessary for these solutions to exist. In \cite{Brihaye:2020vce} it was demonstrated that not only one, but up to three branches of charged black holes with scalar hair can exist in this model. Also charged boson star solutions can be constructed in this model
and these are what we are concerned with here. These have been investigated before in \cite{Brihaye:2021mqk}. Here we will point out that in some limits, these solutions have the features of Frozen stars.

In the second part of this paper, we will discuss the same model, but will add a Horndeski vector-tensor term \cite{Horndeski:1974wa, Horndeski:1976gi} which represents a unique non-minimal coupling between the vector field and curvature. The resulting spherically symmetric solutions are deformed Reissner–Nordstr\"om black holes, as first studied in \cite{muller} and later analyzed in detail in \cite{Verbin:2020fzk}.
We will discuss charged boson stars in this model and demonstrate that frozen states exist within this model and have similar features to those observed in models with non-linear electrodynamics. In particular, we will demonstrate that the scalar self-interaction is not necessary in this case.

\section{The model }
The  model which we will study in the following is defined by the following $4$-dimensional action~:
\begin{equation}
{\cal S}= \int {\rm d}^4 x \sqrt{-g} {\cal L}
\end{equation}
with Lagrangian density given by
\be 
\label{action}
{\cal{L}} = \frac{1}{16 \pi G} R - D^{\mu} \Psi^{\dagger}  D_{\mu} \Psi - U(|\Psi|) - \frac{1}{4} F^{\mu \nu} F_{\mu \nu} + \gamma {\cal{L}}_{VT} \ .
\ee
$R$ is the Ricci scalar, $\Psi$ is a complex-valued scalar field with potential $U(|\Psi|)$
and $F_{\mu\nu}=\partial_{\mu} A_{\nu} - \partial_{\nu} A_{\mu}$ is the field strength tensor of a U(1) gauge field $A_{\mu}$. The scalar field is minimally coupled to this gauge field through the covariant derivative $D_{\mu} = \partial_{\mu} + i q A_{\mu}$.
$G$ is Newton's constant and $q$ the gauge coupling. 
We also allow for a non-minimal coupling between gravity and the U(1) gauge field by introducing a vector-tensor Horndeski term of the form~:
\be
{\cal{L}}_{VT} 
= -\frac{1}{4} \left(F_{\mu \nu} F^{\kappa \lambda} R^{\mu \nu}_{\phantom{\mu \nu} \kappa \lambda}
                          - 4 F_{\mu \kappa} F^{\nu \kappa} R^{\mu}_{{\phantom \mu} \nu}
													+ F_{\mu \nu} F^{\mu \nu} R \right) \ .
\ee
For $\gamma=0$ we recover standard Einstein gravity. In the following, we will demonstrate that both the potential as well as the Horndeski term lead to the existence of so-called {\it frozen stars}. For the potential $U(\vert\Psi\vert)$ we will choose
a potential motivated from gauged supersymmetry breaking \cite{Copeland:2009as}. This reads
	\bea
	\label{eq:susy_pot}
	    U(\vert\Psi\vert) &=& \mu^2 \eta^2 \left[1 - \exp\left(-\frac{\vert\Psi\vert^2}{\eta^2}\right)\right] \ ,
	\eea
    where $\mu$ is the mass of the scalar field and $\eta$ is an energy scale.
Note that this potential behaves like a mass potential for $|\Psi| \ll 1$ and is bounded for $|\Psi| \gg 1$.

In the following, we will discuss stationary, spherically symmetric solutions to the resulting field equations and hence choose the following Ansatz for the metric and matter fields~:
\begin{eqnarray}
& & {\rm d}s^2 = -(\sigma(r))^2 N(r) {\rm d}t^2 + \frac{1}{N(r)} {\rm d}r^2 + r^2\left({\rm d}\theta^2 + \sin^2 \theta {\rm d}\varphi^2 \right) \ \ , \ \  N(r) = 1 - \frac{2m(r)}{r}
\nonumber\\
& & 
A_{\mu} {\rm d} x^{\mu} = V(r) {\rm d} t  \ \ , \ \ 
\Psi=\psi(r) \exp(-i\omega t) \ .
\label{eq:ansatz}
\end{eqnarray} 

We will use dimensionless coordinates and functions and rescale
$r\rightarrow r/\mu$, $\omega \rightarrow \mu \omega$, $\gamma \rightarrow \gamma/\mu^2$, $\psi \rightarrow \eta \psi$, $m(r)\rightarrow m(r)/\mu$. Inserting the Ansatz (\ref{eq:ansatz}) into the field equations, we find a set of four coupled, non-linear differential equations. The gravity equations read~:
\bea
    m' &=& \alpha r^2 \biggl[ \frac{V'^2}{2 \sigma^2} + N \psi'^2 + U(\psi) + \frac{((\omega-q V) \psi)^2}{N \sigma^2} \biggr] +  \alpha \gamma \frac{(1-N)(V')^2}{\sigma^2} \ , \\
		\sigma' &=&  2 \alpha r \sigma \biggl[ \psi'^2 + \frac{((\omega-q V) \psi)^2}{N^2 \sigma^2} \biggr] + 2\alpha \gamma \frac{(V')^2}{r \sigma} \ \ , 
\label{equations_gr}
\eea
where $\alpha=4\pi G \eta^2$. The prime now and in the following denotes the derivative with respect to $r$. For the matter fields we have~:
 \be
	\left(1 + \frac{2 \gamma(1-N)}{r^2} \right)V'' 
	+  \biggl[ \frac{2}{r} -\frac{\sigma'}{\sigma} - \frac{2 \gamma}{r^2} \left((1-N) \frac{\sigma'}{\sigma} + N' \right) \biggr] V' 
		+ \frac{2 q(\omega - q V) \psi^2}{N} = 0 \  ,
\label{equation_1}
		\ee
		\be
			\psi'' +  \left(\frac{2}{r} + \frac{N'}{N} +\frac{\sigma'}{\sigma}\right) \psi' + \frac{(\omega -q  V)^2 \psi}{N^2 \sigma^2} - \frac{1}{2N}
			\frac{dU}{d \psi} = 0  \ .
\label{equation_2}
\ee
Due to the U(1) gauge symmetry, the field equations  depend only on the combination $\omega - q V(r)$. These equations have to be solved subject to the appropriate boundary conditions. To ensure regularity at the origin, we require
\begin{equation}
N(0)=0 \ \ ,  \ \  \psi'(0)=0 \ \ , \ \  V'(0)=0 \ .
\end{equation}
Demanding asymptotic flatness and finite energy leads to
\begin{equation}
\sigma(r\rightarrow \infty) \rightarrow 1  \ \ , \ \ \psi(r\rightarrow \infty) \rightarrow 0 \ \ , \ \ V(r\rightarrow \infty) \rightarrow V_{\infty} \ , 
\end{equation}
with $V_{\infty}$ a real-valued constant. In the following, we will additionally
choose $\psi(0)=C$ with $C$ a real-valued constant. This is to avoid the trivial
solution $\psi(r)\equiv 0$. This choice fixes the value of $\omega$ which no longer
is a free parameter. The scalar field has the following decay
\begin{equation}
\psi(r\rightarrow \infty) \rightarrow \frac{1}{r}\exp\left(- \sqrt{1- (\omega - qV_{\infty})^2} r\right) \ .
\end{equation}
In order to ensure localisation of the scalar field we need to require that
$\omega - q V_{\infty} < 1$. In the following, we will define the parameter $\Omega:=1-(\omega - qV_{\infty})$ and demonstrate that solutions cease to exist when $\Omega \rightarrow 0$.
\\
The physical parameters of the solutions are the (dimensionless) mass $M$, the (dimensionless) electric charge $Q$ and the
(dimensionless) Noether charge $Q_N$.  These can be read off from the metric and matter field functions at infinity
\begin{equation}
\label{eq:infty}
N(r \gg 1)=1-\frac{2M}{r} + \frac{\alpha Q^2}{r^2} + ..... \ \ , \ \ 
V(r)=V_{\infty}  - \frac{Q}{r} + ....
\end{equation}
and the following integral of the $t$-component of the locally conserved Noether current~:
\begin{equation}
\label{eq:noether}
Q_N=\int\limits_0^{\infty} {\rm d} r \ \frac{2r^2 q V\psi^2}{N\sigma}  \ .
\end{equation}
This globally conserved quantity can be interpreted as the number of scalar bosons and 
hence $qQ_N\equiv Q$. 

Note that the coupled differential equations do not possess a globally regular solution for
$\psi\equiv 0$. Additionally setting $\gamma=0$, the system has a closed form solution, the Reissner-Nordstr\"om (RN) solution~:
\begin{equation}
N= 1- \frac{2M}{r} + \frac{2\alpha Q^2}{r^2} \ \  , \ \ \sigma\equiv 1 \ \ , \ \
V(r)=V_{\infty} - \frac{Q}{r} \ 
\end{equation}
with event horizon at $r_{+}=M(1+\sqrt{1-2\alpha Q^2/M^2})$. The extremal RN solution has
$r_+=M=\sqrt{2\alpha}Q$. 
For $\gamma\neq 0$ and $\psi\equiv 0$ the solutions were studied in \cite{Verbin:2020fzk}.

\section{Numerical results}
\subsection{$\gamma=0$: the role of scalar self-interaction and charge}
We first discuss the solutions obtained in the case of a minimal coupling to gravity i.e. for $\gamma=0$. In order to show that the self-interaction term is crucial for the existence of frozen stars, we first discuss the case of a simple mass potential for the scalar field, i.e. we choose $U(\psi)=\psi^2$. In this case,  $\alpha$ can be set to unity without loss of generality by applying a rescaling of the matter fields $V\rightarrow V/\alpha$, $\psi\rightarrow \psi/\alpha$.  Hence, the only remaining coupling is $q$ that gets rescaled as $q\rightarrow \alpha q$.

\begin{figure}[h!]
{\includegraphics[width=8cm]{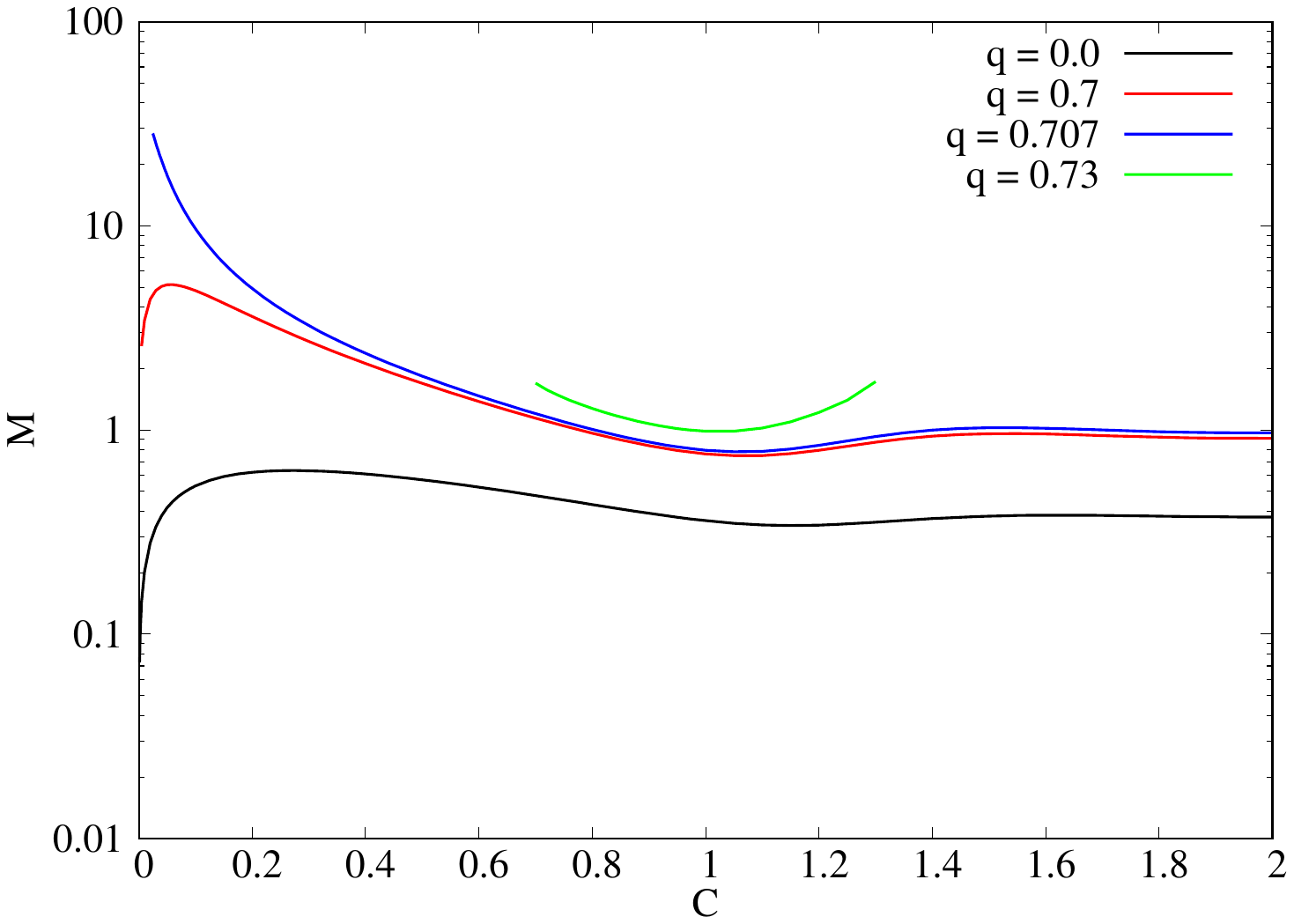}}
{\includegraphics[width=8cm]{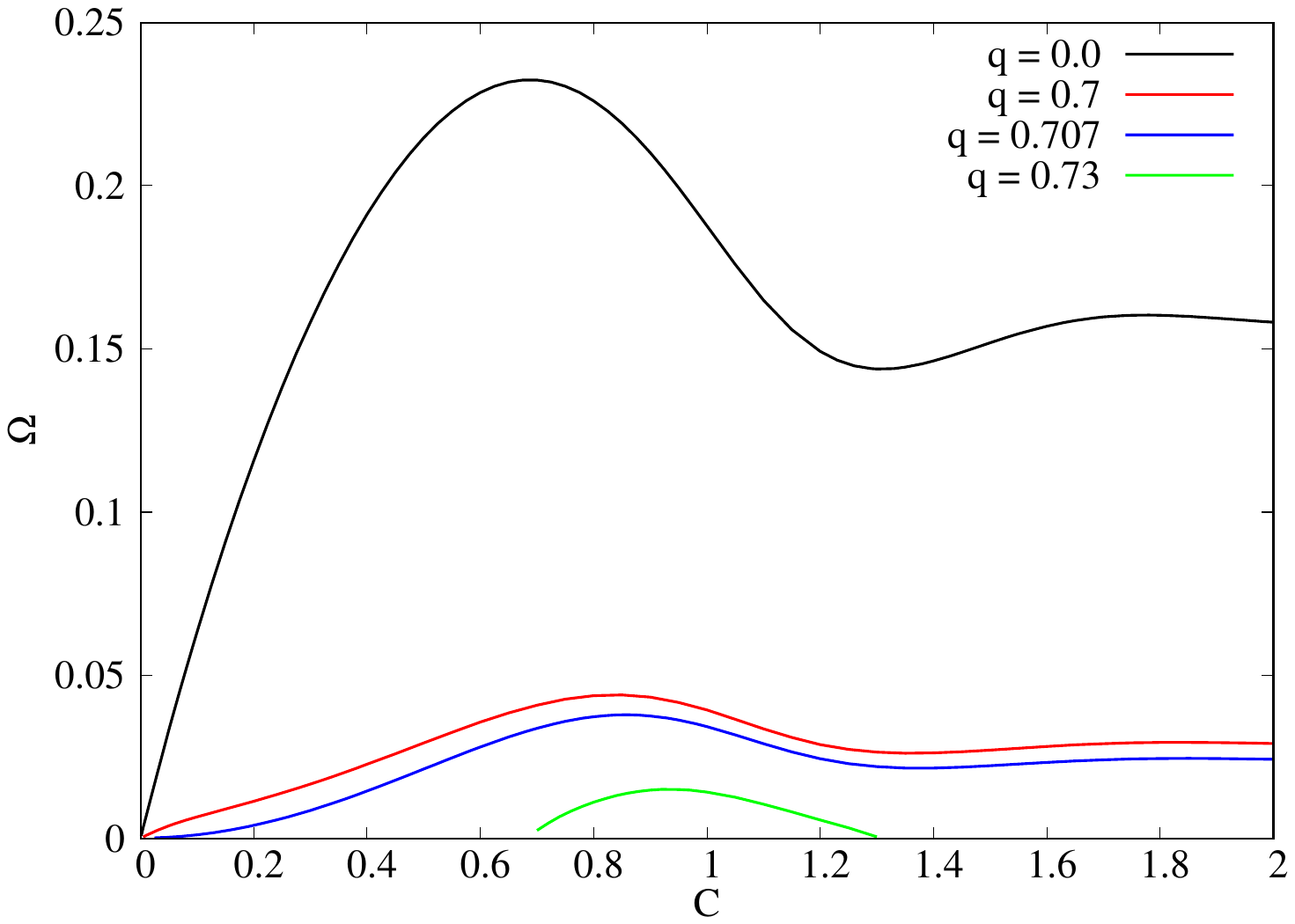}}
\caption{{\it Left}: We show the dependence of the mass $M$ on the value $C$ of the scalar field $\psi(r)$ at the origin for different values of $q$.
Right: The  corresponding value of $\Omega=1 - (\omega - qV_{\infty})^2$.
\label{fig_1}
}

\end{figure}

\begin{figure}[h!]
\begin{center}
{\includegraphics[width=8cm]{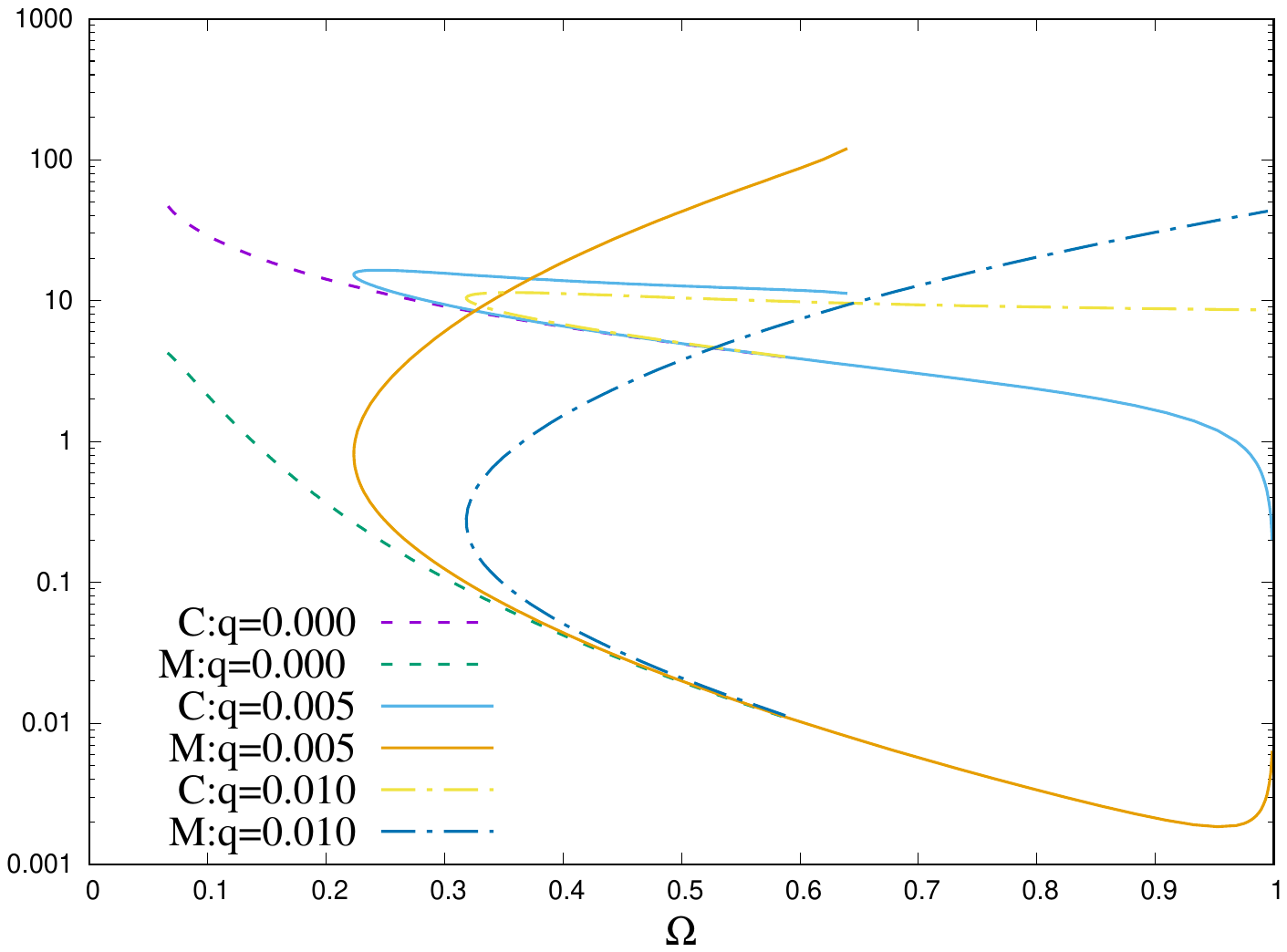}}
{\includegraphics[width=8cm]{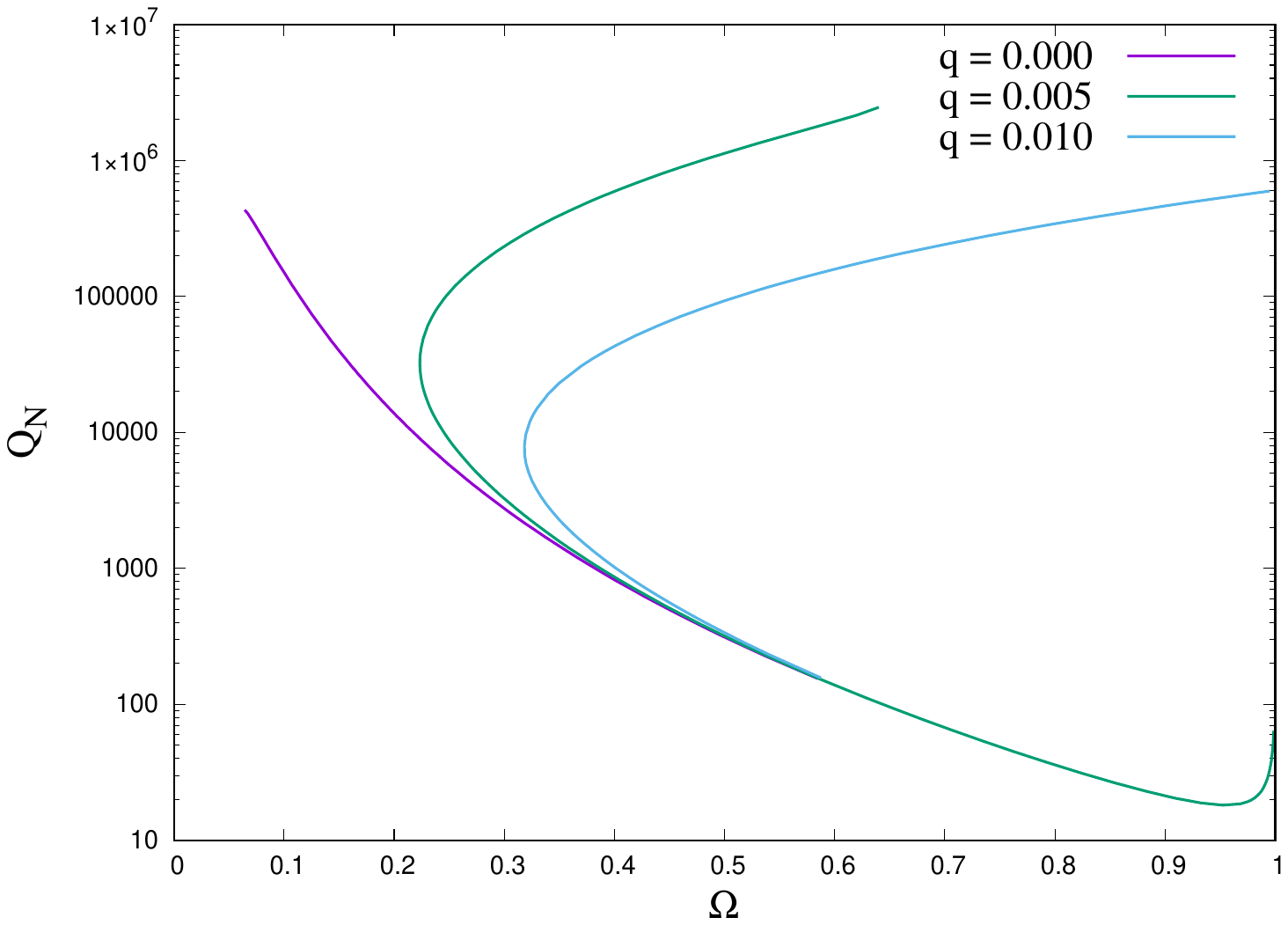}}
\caption{{\it Left}: The show the value $C$ of the scalar field $\psi(r)$ at the origin as well as the mass $M$ of the solutions in dependence on $\Omega$ for $\alpha=0.0001$
and several values of $q$. {\it Right}: We show the dependence of the Noether charge $Q_N$ on $\Omega$ for the same solutions.
\label{fig_2}
}
\end{center}
\end{figure}

These solutions have been studied in \cite{Pugliese:2013gsa} and revisited recently in \cite{Lopez:2023phk}. In order to be able to compare with the self-interacting case, we
put the emphasis here on the domain of existence of solutions in the $q$-$C$-plane. 
A family  of boson stars corresponding to a specific value of the gauge coupling constant $q$ can be  
constructed numerically by increasing gradually the parameter $C$.  Increasing $C$,  the value of the function $\sigma(r)$ at the origin, $\sigma(0)$, decreases from unity. Our numerical results strongly suggest that $\sigma(0)$ approaches zero for some maximal value of $C$. 
The dependence of the mass $M$ and the parameter $\Omega$ on the central value $C$ is shown in  Fig. \ref{fig_1}.
As described in  \cite{Lopez:2023phk}, for $q > 1/ \sqrt 2$ we observe that solutions do no longer exist for arbitrary $C$, but only on a very limited interval. Increasing $q$ beyond this value decreases this interval until solutions cease to exist for $q > 0.75$  \cite{Brihaye:2023hwg}. For all solutions constructed, we find that the metric function $N(r)$ presents a minimum
that stays finite and hence never approaches zero. This suggests that {\it charged boson stars in standard Einstein gravity with a simple mass potential cannot reach a frozen state.} However, this changes when we introduce the self-interacting potential (\ref{eq:susy_pot}) as we will demonstrate below. Since we cannot perform a rescaling similar to the one used in the mass potential case, $\alpha$ and $q$ are now free coupling constants. In the following, we will set $\alpha=0.0001$. For $q=0$
 the solutions can be characterized by $\Omega$ and the parameter $C$, see Fig. \ref{fig_2}. In this figure, we show the mass $M$ and the parameter $C$ in dependence of $\Omega$ for several values of $q$ including $q=0$. Solutions with large $C$
do exist, the metric function $N(r)$ possesses a local minimum that, however, stays perfectly finite along the branches. Typically we find that the minimum of $N$ is $\sim 0.5$. 

\begin{figure}[h!]
\begin{center}
\includegraphics[width=8cm]{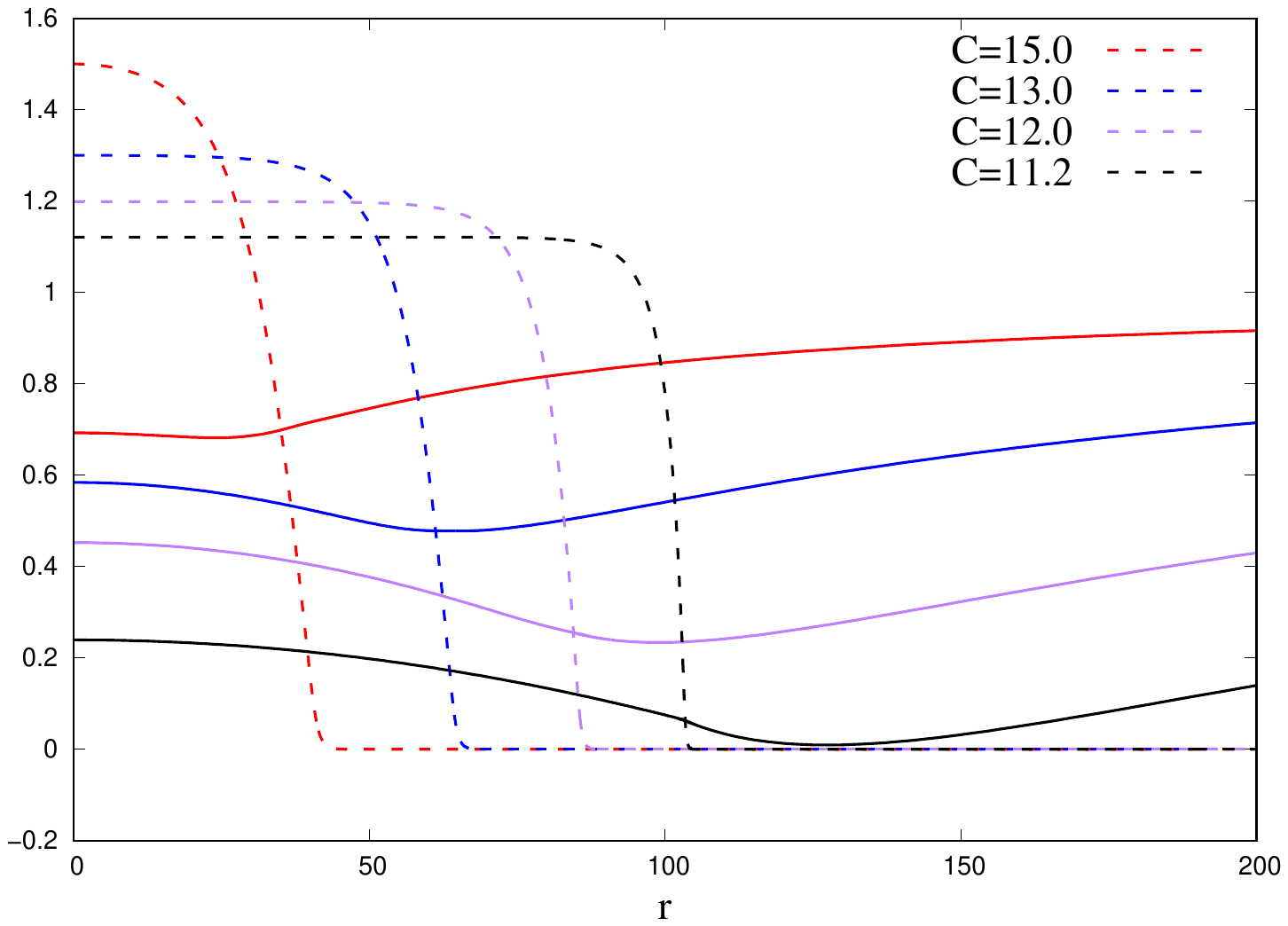}
\includegraphics[width=8.2cm]{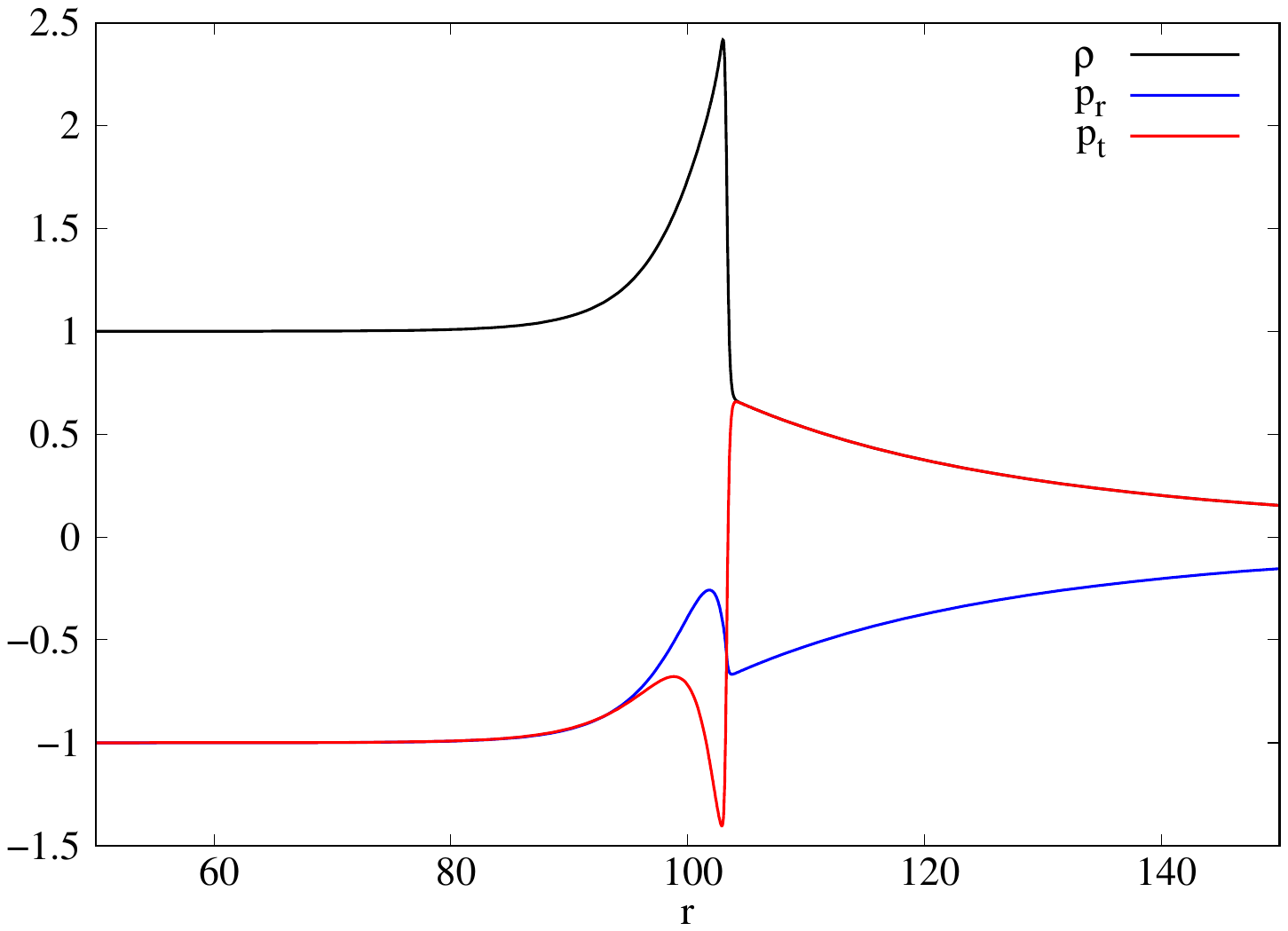}
\caption{{\it Left:} We show the profiles  of the metric function $-g_{tt}=\sigma^2 N$ (solid) and of the scalar field function $\psi(r)$ (dashed) for three charged boson star solutions on the branch $B$. Here $\alpha=0.0001$, $q=0.005$ and the three solutions have different values of $C$, i.e. different values of the scalar field at the origin. Clearly, we see that the metric function approaches a double zero at some intermediate value of the radial coordinate $r$ when decreasing $C$. {\it Right:} The corresponding energy density $\rho$, radial pressure $p_r$ and transversal pressure $p_t$ for the solution with $C=11.2$.
\label{fig:profile_1}
}
\end{center}
\end{figure}

This changes when $q > 0$. As can be clearly seen in Fig. \ref{fig_2}, we find two branches of solutions in $\Omega$. These two branches join at a minimal value of $\Omega=\Omega_{min}$. This minimal value depends on $q$, e.g. 
$\Omega_{min} \approx 0.223$ for $q = 0.005$, and $\Omega_{min} \approx 0.318$ for $q = 0.01$. In the following, we will refer to the branch of lower $M$ as branch A, while the branch of higher $M$ will be denoted branch B. 
When $q$ is sufficiently large (typically $q \sim 0.01$) the two branches 
can be extended to $\Omega = 1$ and the local minimum of the metric function $N(r)$, again, is well above zero. 
Interestingly, for smaller values of $q$ (typically $q \sim 0.005$) the branch $B$ stops at some intermediate 
value of $\Omega=\Omega_c$ with $\Omega_{min} < \Omega_c < 1$. We find that the branch terminates here because the minimum of $N(r)$ approaches zero at some value $r_c$ of the radial coordinate. This approach is shown in Fig. \ref{fig:profile_1} (left). Clearly, the solution develops a double zero in this case. These solutions have previously been
discussed in the context of forming a thin shell between the boson star interior and its exterior \cite{Brihaye:2021mqk}. This behaviour can be understood as follows~: 
for $q=0$, an arbitrary number $Q_N$ of scalar quanta can coexist bounded by the gravitational force, which is no longer true when $q > 0$. In this case, the quanta
also carry electric charge and this leads to an electric repulsion. The balance between the gravitational force and the electric repulsion 
allows for compact boson stars to exist, so dense that a horizon can be formed. For very large values of $q$, the electric repulsion is too large to allow very compact objects.

In order to make the connection to the Mazur-Mottola solution \cite{Mazur:2001fv}, we check the components of the energy-momentum tensor for the solutions.
The energy density $\rho=-T^t_t$, radial pressure $p_r=T^r_r$ and transverse pressure $p_{t}=T_{\theta}^{\theta}=T^{\varphi}_{\varphi}$ are given by
\begin{align}
\label{eq:density_pressure}
\rho = \frac{V'^2}{2\sigma^2}  + N {{\psi}}'^2 + \frac{(\omega - q V)^2 {{\psi}}^2}{N \sigma^2} +  U({\psi})  \ , \\
p_r =  - \frac{V'^2}{2\sigma^2}  +  N {{\psi}}'^2 + \frac{(\omega - q V)^2 {{\psi}}^2}{N \sigma^2} - U({{\psi}}) \ , \\
p_{t}=\frac{V'^2}{2\sigma^2}  - N {{\psi}}'^2 +\frac{(\omega - q V)^2 {{\psi}}^2}{N \sigma^2} - U({{\psi}})   \ .
\end{align}
Clearly, $\rho \geq p_r$ and
$\rho \geq p_{t}$, i.e. the model fulfills the causality bound as well as the weak energy condition. In \cite{Mazur:2001fv} it was argued that in a model with anisotropic fluid that fulfills  $\rho=-p_r$ so-called {\it frozen stars} can be constructed. In Fig. \ref{fig:profile_1} (right) we show the energy-momentum tensor components of our 
solution with $C=11.2$. Clearly, the interior of the solution is free from singularities
and $\rho=-p_r=-p_t$, i.e. it possesses a de Sitter core. There exists a spherical shell of finite thickness that interpolates with the RN exterior, where $\rho=p_t=-p_r$. Hence, this configuration is globally regular interpolating between a regular interior
and a RN exterior with the event horizon replaced by a shell of finite thickness. 

\subsection{$\gamma\neq 0$}
In the following, we would like to demonstrate that the frozen states of boson stars
also appear without self-interaction of the scalar field (which is necessary in standard Einstein gravity, see above), but when considering vector-tensor Horndeski gravity.
In the following, we will replace the self-interaction potential by a mass term for the
scalar field, i.e. $U(\psi)=\psi^2$.

\begin{figure}[h!]
\begin{center}
{\includegraphics[width=8cm]{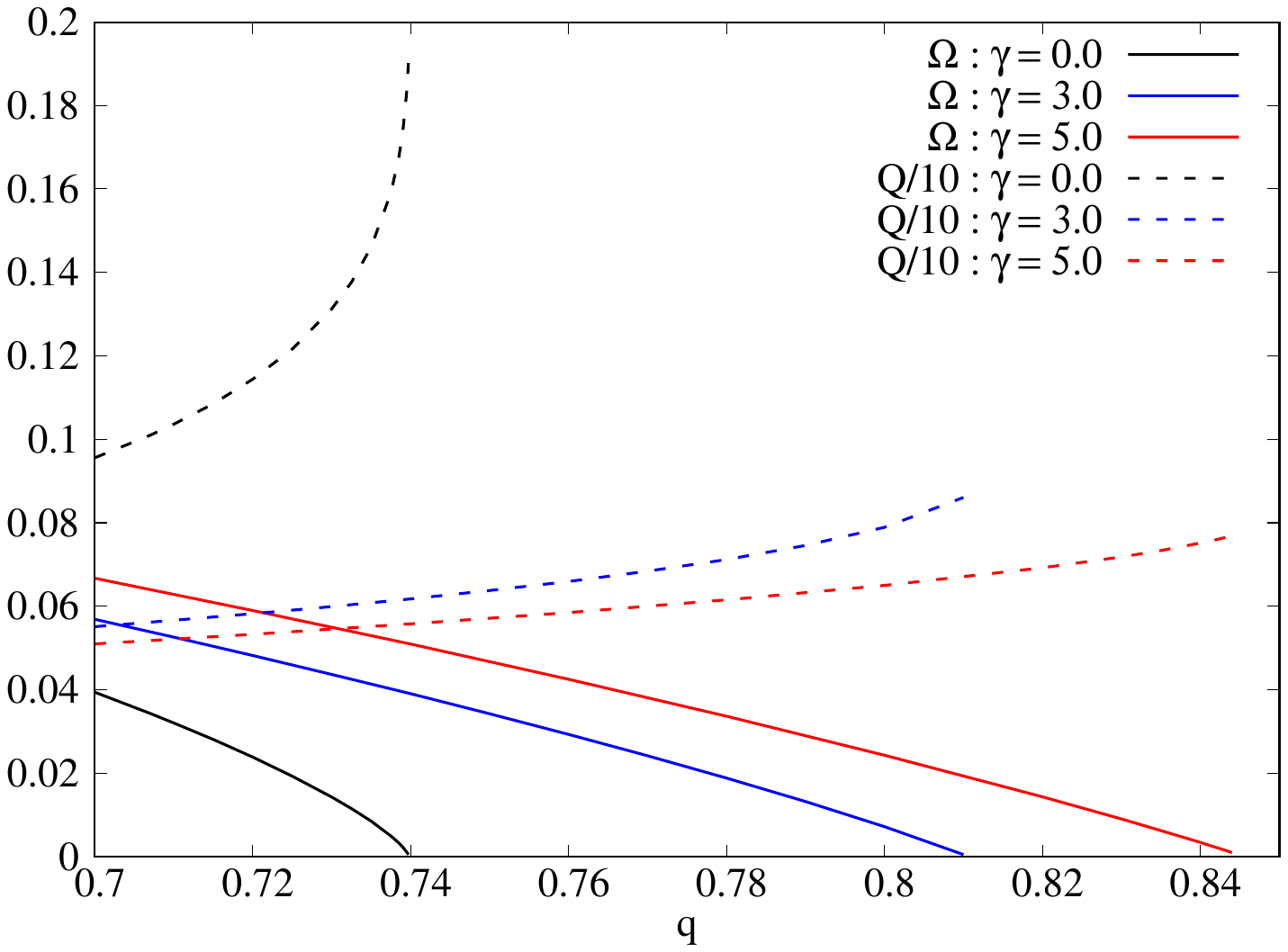}}
{\includegraphics[width=8cm]{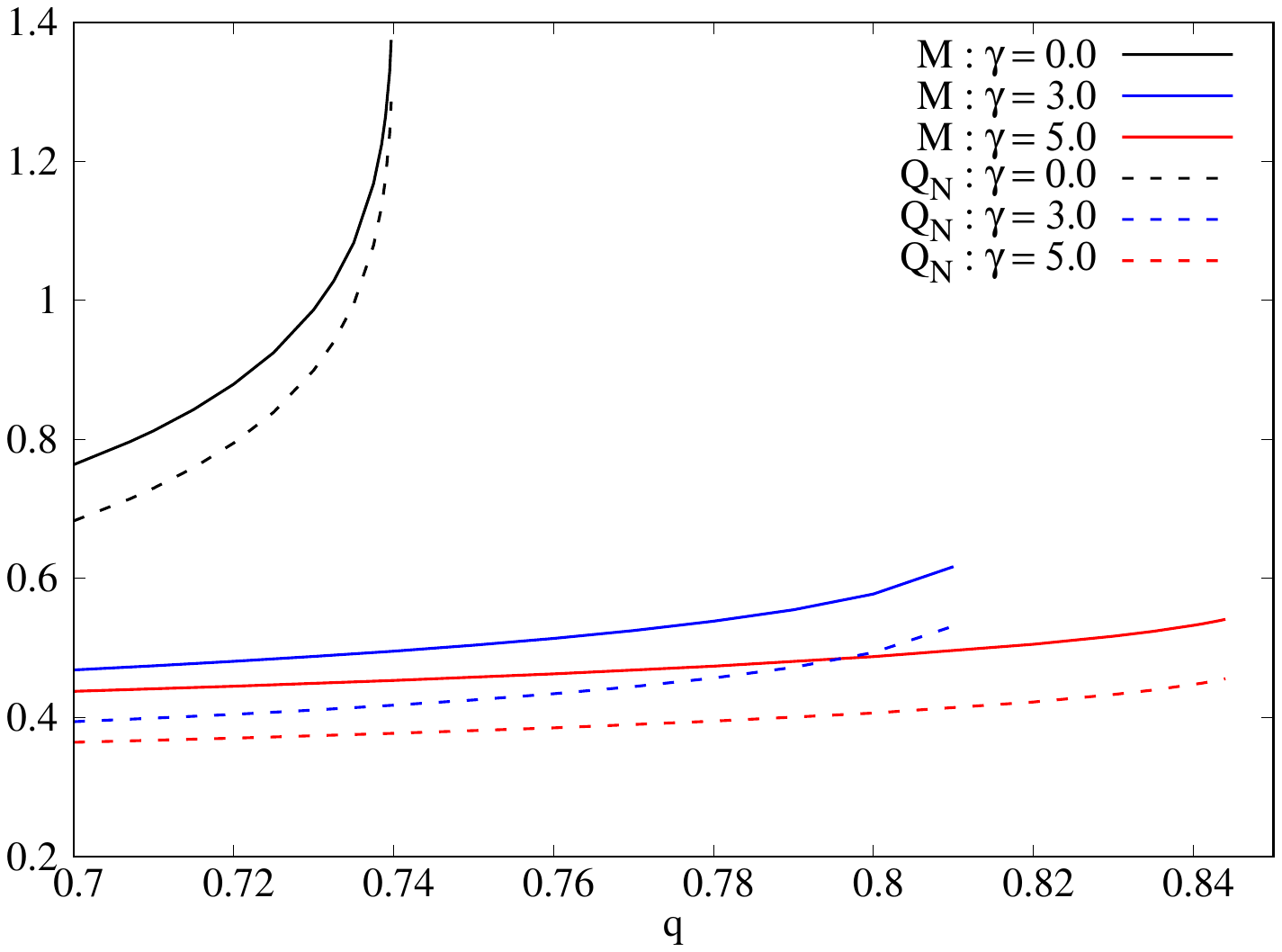}}
\caption{{\it Left}: We show the dependence of 
$\Omega$ and of the electric charge $Q$ on the gauge coupling constant $q$ for constant central value of the scalar field, $\psi(0)=1$,
and three different values of $\gamma$. {\it Right}: We show the corresponding mass $M$ and Noether charge $Q_N$ of the Horndeski vector-tensor boson star in dependence on $q$.
\label{fig:horndeski}
}
\end{center}
\end{figure}

Our numerical results indicate that for $\gamma > 0$, the properties of the solutions are qualitatively similar to those discussed above. 
Quantitatively, we observe that bosons stars with positive $\gamma$ exist generically 
for higher values  of the gauge coupling $q$ as compared to the $\gamma=0$ case. 
This is shown in Fig. \ref{fig:horndeski}, where we give the dependence of $\Omega$ and $Q$
on the gauge coupling $q$ for three different values of $\gamma$ including $\gamma=0$ (left) for a fixed central value of the scalar field $\psi(0)=1$.
The limit $\Omega \rightarrow 0$ is reached for $q=q_{\rm max}$. We find that $q_{\rm max}$ increases with $\gamma$. The dependence of the mass $M$ and the Noether charge $Q_N$ on $q$ also changes with $\gamma$, see Fig. \ref{fig:horndeski} (right). For $\gamma=0$ both $M$ and $Q_N$ increase quicker with $q$ than in the $\gamma > 0$ case. Hence, vector-tensor Horndeski boson stars have lower mass and Noether charge for the same value of $q$ as compared to the 
boson stars in standard Einstein gravity.

\begin{figure}[h!]
\begin{center}
{\label{c1}\includegraphics[width=9cm]{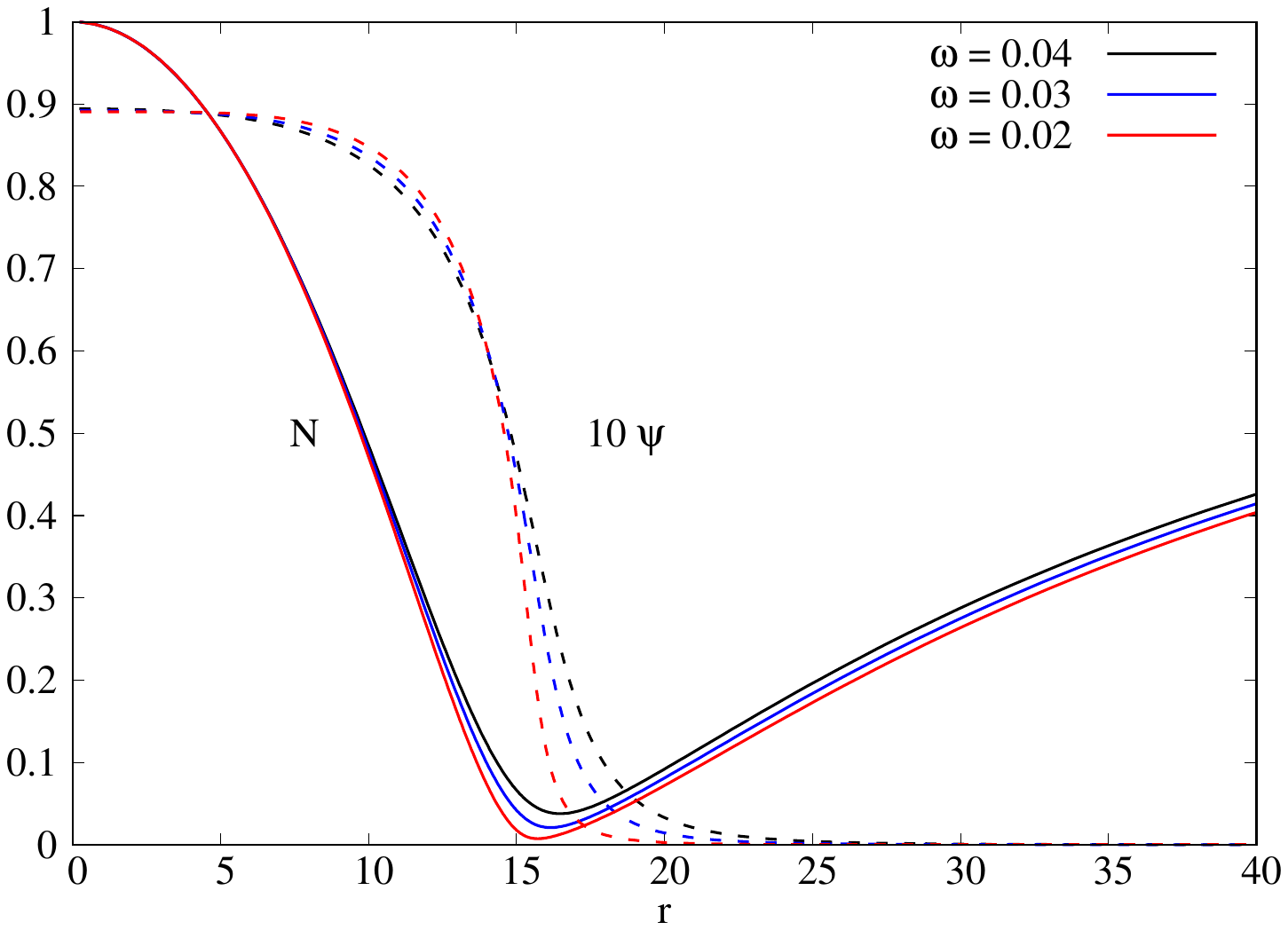}}
\caption{We show the profiles of the metric function $N(r)$ and of the scalar field function $\psi(r)$ for a vector-tensor Horndeski boson star with $q=0.7$, $\gamma=-3$ and three values of $\omega$.
\label{fig:horndeski2}
}
\end{center}
\end{figure}

Interestingly, choosing $\gamma < 0$ leads to new features. In the following, we will demonstrate this for the case $q=0.7$, $\gamma=-3$. Let us first note that
vector-tensor boson stars do exist in the limit $C \to 0$ which corresponds to $\Omega \rightarrow 0$.
However, decreasing the frequency parameter $\omega$ we find that the solution 
approaches a configuration that looks like a frozen star. This configuration possesses a quasi-horizon with the metric function $N(r)$ coming close to zero at some intermediate value of the radial coordinate. This is shown in Fig. \ref{fig:horndeski2}, where we give the profiles of the metric function $N(r)$ together with the profile of the scalar field function $\psi(r)$
for different values of $\omega$. Decreasing $\omega$ (which corresponds to increasing $\Omega$ in this case), the value of the minimum of $N(r)$ approaches zero, while the scalar field is well-behaved.

\begin{figure}[h!]
\begin{center}
{\includegraphics[width=8cm]{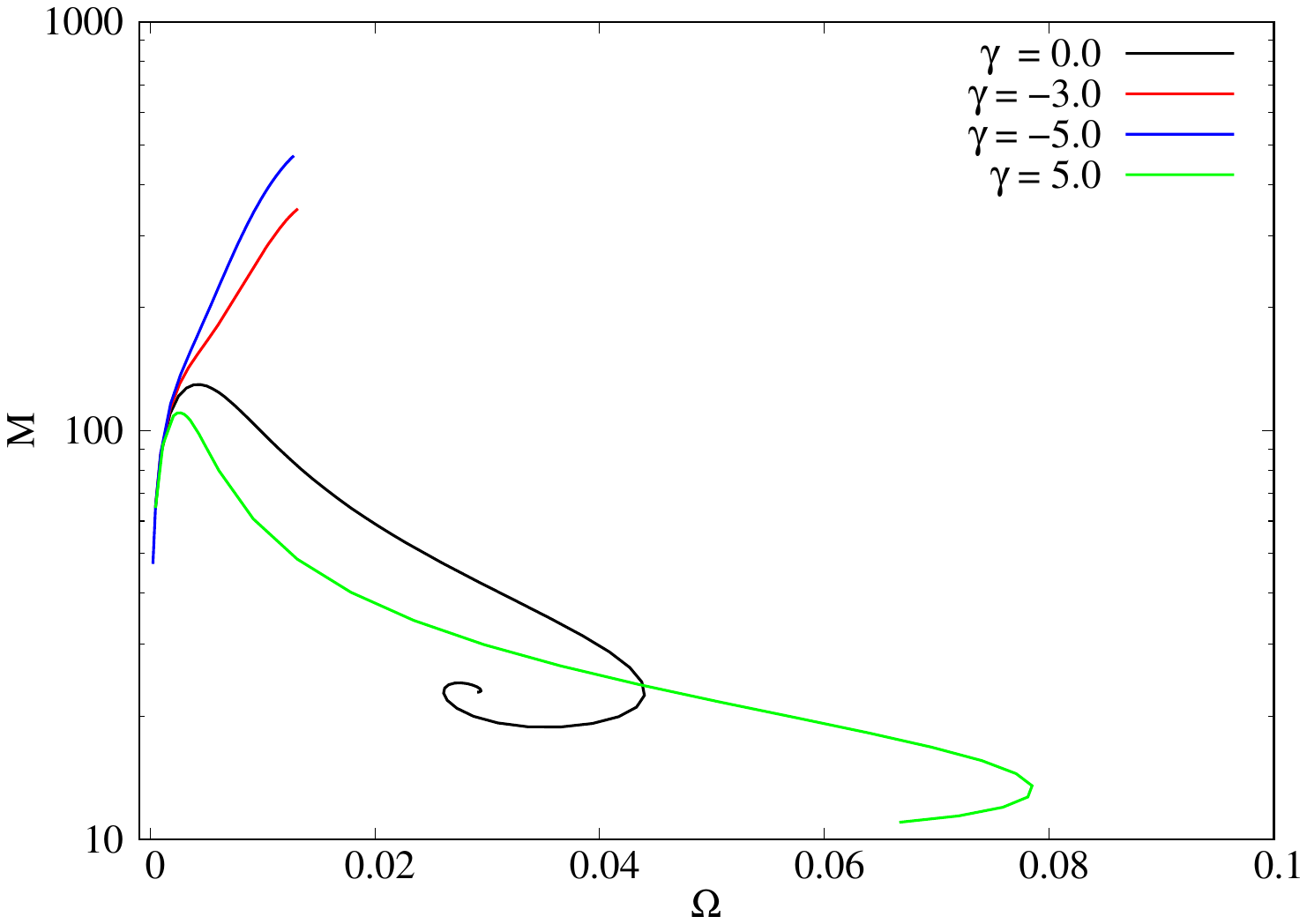}}
{\includegraphics[width=8cm]{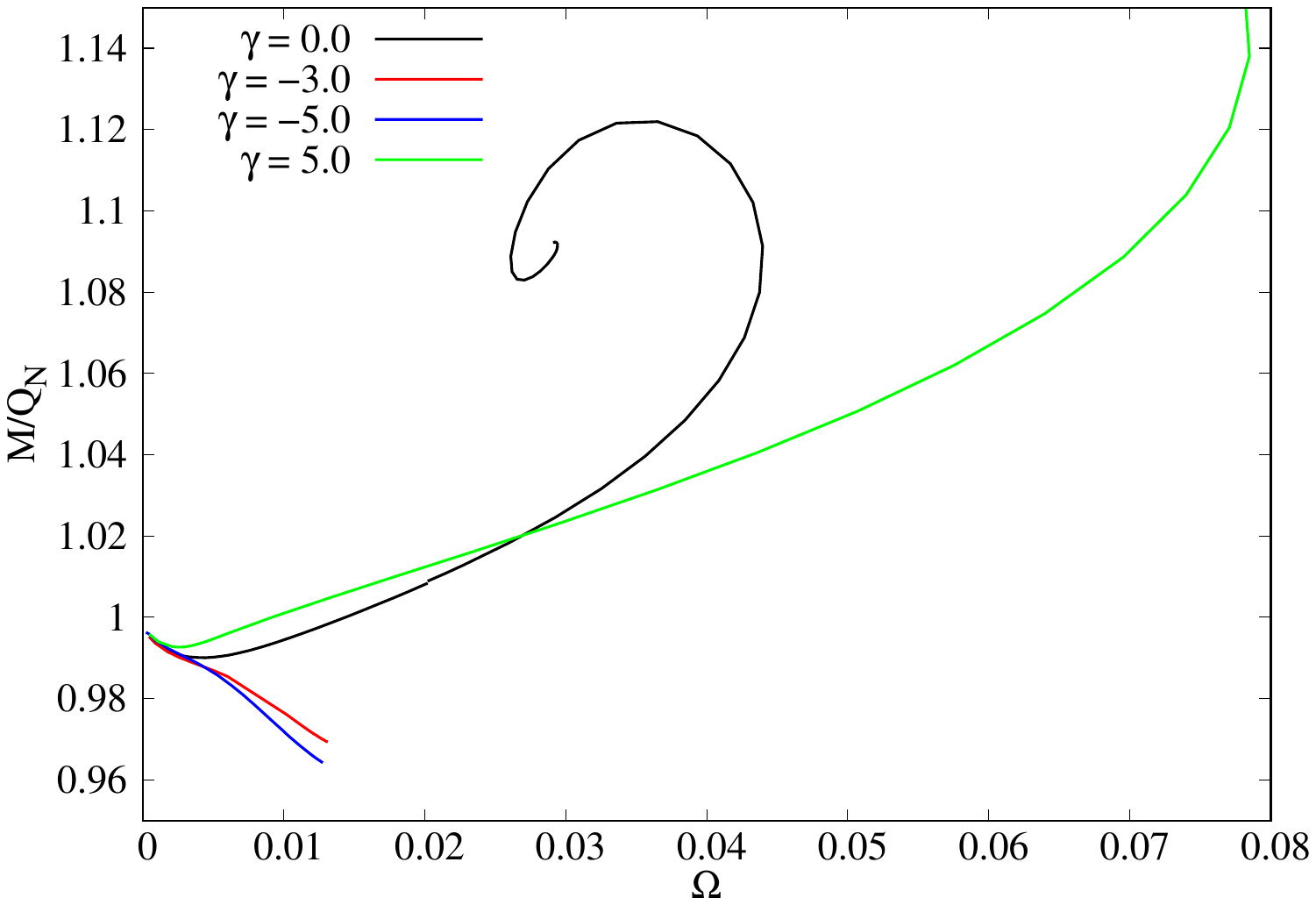}}
\caption{{\it Left}: We show the dependence of the mass $M$ on the parameter 
$\Omega$ for $q=0.7$ and different values of $\gamma$.
{\it Right}: We show the ratio $M/Q_N$ in dependence on $\Omega$ for the same solutions.
\label{fig:horndeski3}.
}
\end{center}
\end{figure}

In general, we observe that this limiting behaviour leads to the fact that the branches of solutions with $\gamma < 0$ are very limited in both $C$ and $\Omega$. This can be seen in Fig. \ref{fig:horndeski3}, where we compare the dependence of the mass $M$ (left) and the
ratio between the mass $M$ and the Noether charge $Q_N$ (right) on $\Omega$.
For $\gamma \geq 0$, we find that more than one branch of solutions exists, while for $\gamma < 0$ only one branch appears. In this latter case, the branches end exactly due to the limiting behaviour described above, i.e. the branches end at a frozen state. To be able to compare this, we note that the values of $\omega=0.04$, $0.03$ and $0.02$ correspond to 
$\Omega=0.01314$, $0.01336$, and $0.01360$, respectively. It is clear from our numerical data is that the frozen states of Horndeski vector-tensor boson stars are larger in mass $M$ than their non-frozen counterparts. However, we find that $M/Q_N < 1$ for all solutions with $\gamma < 0$. This means that these configurations are stable with respect to the decay into $Q_N$ individual bosons of mass $\mu=1$.

\section{Discussion}

\begin{figure}[h!]
\begin{center}
{\includegraphics[width=8cm]{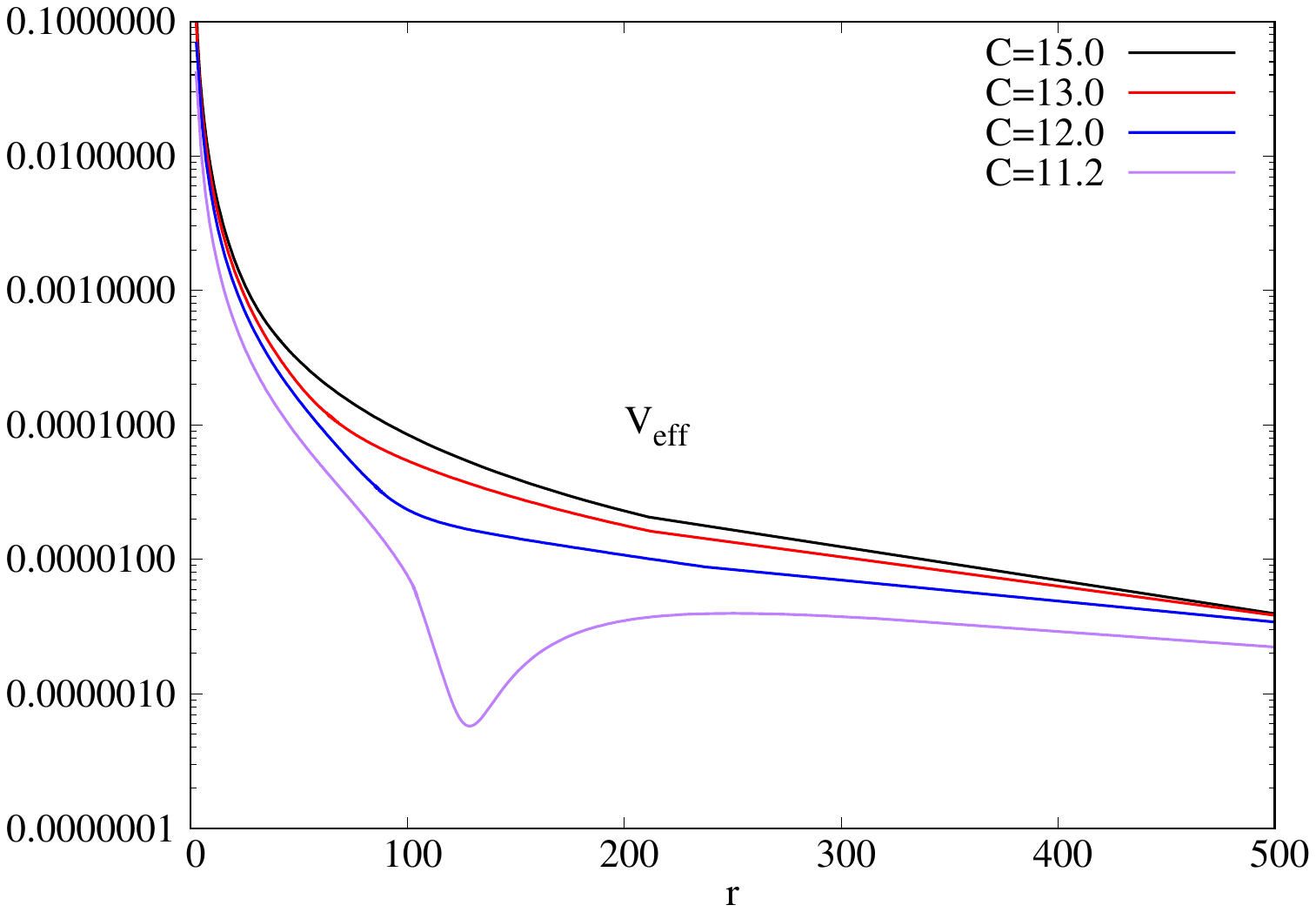}}
{\includegraphics[width=8cm]{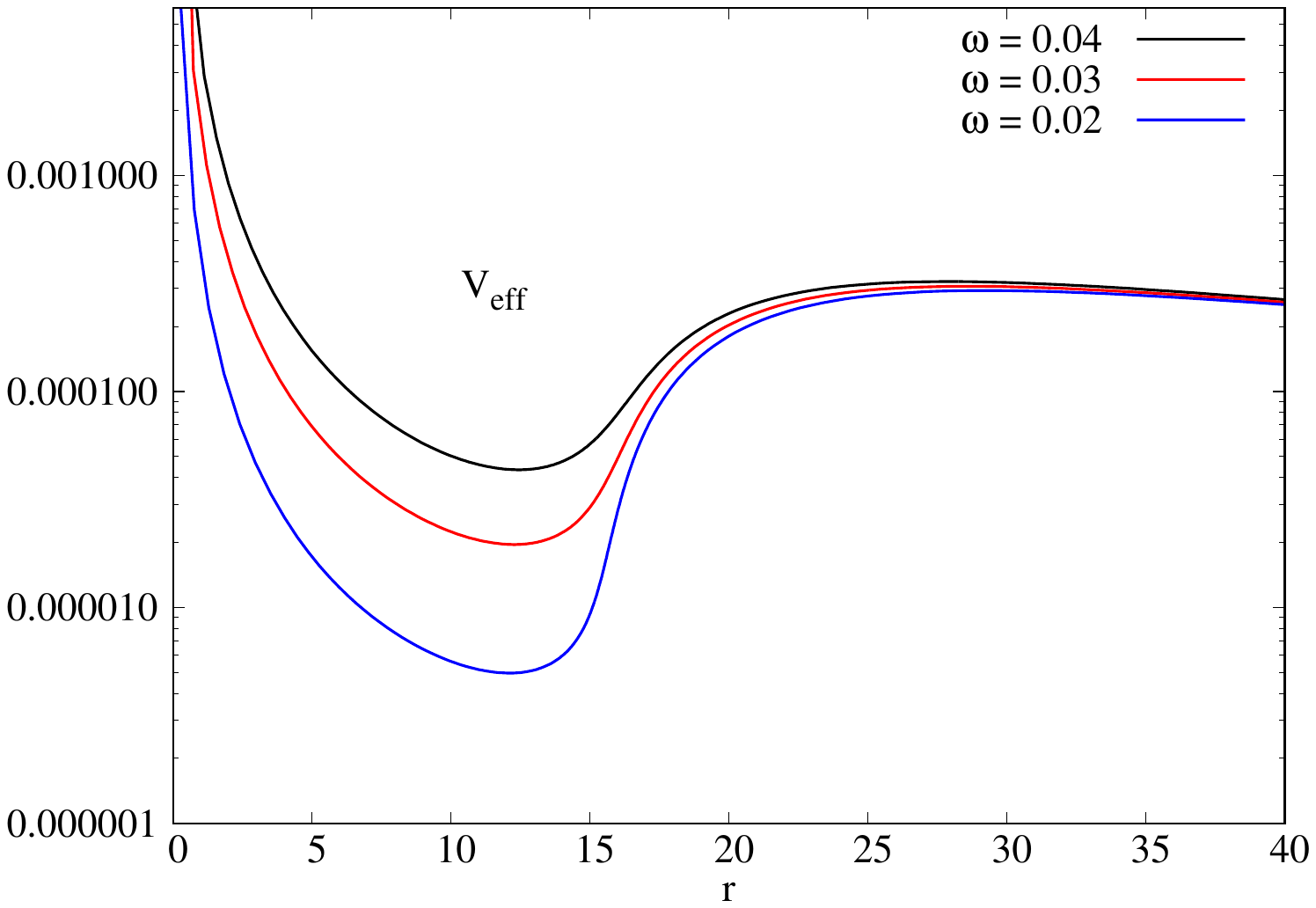}}
\caption{{\it Left}: We show the effective potential for light-like particle motion in the space-time of a charged frozen boson stars. {\it Right}: We show the effective potential for light-like particle motion in the space-time of the Horndeski charged frozen boson stars.
\label{fig:effectivepot}.
}
\end{center}
\end{figure}

In this paper, we have studied the existence of frozen states of charged boson stars.
We find that the standard ungauged scalar field model does not allow for these states.
However, when gauging the U(1) symmetry and introducing a self-interaction for the scalar field, we find that these states appear for intermediate values of the gauge coupling. 
Moreover, adding a vector-tensor Horndeski term, we find that these frozen states also appear without the scalar self-interaction.
In order to understand observational signatures of these configurations, let us consider the (equatorial) geodesic motion of light-like test particles which can be described by
the following equation~:
\begin{equation}
\sigma^2 \dot{r}^2 + V_{\rm eff}(r) = E^2 \ , \ V_{\rm eff}= L_z^2 \frac{N\sigma^2}{r^2}  
\end{equation}
where $E$ is the energy of the test particle, while $L_z$ is its angular momentum with respect to the $z$-axis. The RN solution possess a stable and one unstable lightring, but only the latter can be observed as it lies outside the event horizon. In Fig. \ref{fig:effectivepot}, we show the effective potential at the approach to the frozen state. This demonstrates  that these configurations  possess an inner stable and an outer unstable lightring. The stable lightring lies within the thin, intermediate shell, while the unstable lightring can be found in the black hole exterior, which for $\gamma=0$
corresponds to the RN solution. Possessing a lightring, these objects qualify as ultracompact objects (UCO) and it has been shown that for UCOs  lightrings always come in pairs \cite{Cunha:2017qtt}. The existence of a stable lightring has been suggested to lead to instabilities of UCOs \cite{Cardoso:2014sna}, which has been confirmed by a full numerical simulation in pure scalar field models \cite{Cunha:2022gde}. The frozen states in our model
possess additional gauge fields and, in particular, a de Sitter core. So, it would be interesting to understand whether the results of \cite{Cunha:2022gde} also apply to the frozen states discussed here.

 \end{document}